\input harvmac
\def\half{{1 \over 2}}
\def\dzm{{\partial}}

\def\p{{\partial}}
\def\s{{\sigma}}

\def\pd{{\dot +}}
\def\md{{\dot -}}

\def\a {{\alpha}}
\def\b {{\beta}}
\def\g {{\gamma}}
\def\d {{\delta}}
\def\e {{\epsilon}}

\def\ad {{\dot\alpha}}
\def\bd {{\dot\beta}}

\def\t {{\theta}}
\def\tb {{\theta^*}}

\def\tba {{\theta^{*\ad}}}

\def\tbar {{\bar\theta}}
\Title{\vbox{\hbox{IFT-P.044/2003,}}
       \vbox{\hbox{CALT-68-2455,}}
       \vbox{\hbox{HUTP-03/A064}}   }
{\vbox{\centerline{\bf On the Worldsheet Derivation 
of Large N Dualities for the Superstring}}}
\bigskip\centerline{Nathan Berkovits}
\bigskip\centerline{Instituto de F\'{\i}sica Te\'orica, Universidade Estadual
Paulista}
\centerline{Rua Pamplona 145, S\~ao Paulo, SP 01405-900, BRASIL}
\bigskip\centerline{Hirosi Ooguri}
\bigskip\centerline{California Institute of Technology 452-48}
\centerline{Pasadena, CA 91125, USA}
\bigskip\centerline{ Cumrun Vafa }
\bigskip\centerline{Jefferson Physical Laboratory, Harvard University}
\centerline{Cambridge, MA 02138, USA}
\vskip .2in
Large $N$ topological string dualities have led to a class of proposed
open/closed dualities for superstrings. In the topological
string context, the worldsheet derivation of these dualities has already
been given.
In this paper we take the first step in deriving
the full ten-dimensional superstring dualities by showing how
the dualities arise on the superstring worldsheet
at the level of $F$ terms.
As part of this derivation, we show for $F$-term computations that
the hybrid formalism for the superstring is equivalent to a
$\hat c=5$ topological string in ten-dimensional spacetime.
Using the $\hat c=5$ description, we then show that the D brane
boundary state for the ten-dimensional
open superstring naturally emerges on the worldsheet of
the closed superstring dual.

\Date{October 2003}
%\draft
%\VafaWI
\lref\VafaWI{
C.~Vafa,
``Superstrings and topological strings at large $N$,''
J.\ Math.\ Phys.\  {\bf 42}, 2798 (2001)
[arXiv:hep-th/0008142].
%%CITATION = HEP-TH 0008142;%%
}
\lref\hybrid{N. Berkovits, ``Covariant quantization of the
Green-Schwarz superstring in a Calabi-Yau background,''
Nucl. Phys. {\bf B431}, 258 (1994),
[arXiv: hep-th/9404162].}
%\BershadskyCX
\lref\BershadskyCX{
M.~Bershadsky, S.~Cecotti, H.~Ooguri and C.~Vafa,
``Kodaira-Spencer theory of gravity and exact results
for quantum string amplitudes,''
Commun.\ Math.\ Phys.\  {\bf 165}, 311 (1994)
[arXiv:hep-th/9309140].
%%CITATION = HEP-TH 9309140;%%
}
%\OoguriCK
\lref\OoguriCK{
H.~Ooguri, Y.~Oz and Z.~Yin,
``D-branes on Calabi-Yau spaces and their mirrors,''
Nucl.\ Phys.\ B {\bf 477}, 407 (1996)
[arXiv:hep-th/9606112].
%%CITATION = HEP-TH 9606112;%%
}
%\KlebanovHB

%\LabastidaZP
\lref\LabastidaZP{
J.~M.~Labastida and M.~Marino,
``Polynomial invariants for torus knots and topological strings,''
Commun.\ Math.\ Phys.\  {\bf 217}, 423 (2001)
[arXiv:hep-th/0004196].
%%CITATION = HEP-TH 0004196;%%
}

%\RamadeviGQ
\lref\RamadeviGQ{
P.~Ramadevi and T.~Sarkar,
``On link invariants and topological string amplitudes,''
Nucl.\ Phys.\ B {\bf 600}, 487 (2001)
[arXiv:hep-th/0009188].
%%CITATION = HEP-TH 0009188;%%
}

%\LabastidaTS
\lref\LabastidaTS{
J.~M.~Labastida and M.~Marino,
``A new point of view in the theory of knot and link invariants,''
arXiv:math.qa/0104180.
%%CITATION = MATH-QA 0104180;%%
}

\lref\KlebanovHB{
I.~R.~Klebanov and M.~J.~Strassler,
``Supergravity and a confining gauge theory: Duality cascades and
$\chi SB$-resolution of naked singularities,''
JHEP {\bf 0008}, 052 (2000)
[arXiv:hep-th/0007191].
%%CITATION = HEP-TH 0007191;%%
}
%\CachazoSG
\lref\CachazoSG{
F.~Cachazo, B.~Fiol, K.~A.~Intriligator, S.~Katz and C.~Vafa,
``A geometric unification of dualities,''
Nucl.\ Phys.\ B {\bf 628}, 3 (2002)
[arXiv:hep-th/0110028].
%%CITATION = HEP-TH 0110028;%%
}
%\MaldacenaYY
\lref\MaldacenaYY{
J.~M.~Maldacena and C.~Nunez,
``Towards the large $N$ limit of pure ${\cal N} = 1$ super Yang Mills,''
Phys.\ Rev.\ Lett.\  {\bf 86}, 588 (2001)
[arXiv:hep-th/0008001].
%%CITATION = HEP-TH 0008001;%%
}
 %\WittenFB
\lref\WittenFB{
E.~Witten,
``Chern-Simons gauge theory as a string theory,''
Prog.\ Math.\  {\bf 133}, 637 (1995)
[arXiv:hep-th/9207094].
%%CITATION = HEP-TH 9207094;%%
}
%\BershadskyCX
\lref\BershadskyCX{
M.~Bershadsky, S.~Cecotti, H.~Ooguri and C.~Vafa,
``Kodaira-Spencer theory of gravity and exact results for quantum string amplitudes,''
Commun.\ Math.\ Phys.\  {\bf 165}, 311 (1994)
[arXiv:hep-th/9309140].
%%CITATION = HEP-TH 9309140;%%
}

%\OoguriGX
\lref\OoguriGX{
H.~Ooguri and C.~Vafa,
``Worldsheet derivation of a large $N$ duality,''
Nucl.\ Phys.\ B {\bf 641}, 3 (2002)
[arXiv:hep-th/0205297].
%%CITATION = HEP-TH 0205297;%%
}
%\GopakumarKI
\lref\GopakumarKI{
R.~Gopakumar and C.~Vafa,
``On the gauge theory/geometry correspondence,''
Adv.\ Theor.\ Math.\ Phys.\  {\bf 3}, 1415 (1999)
[arXiv:hep-th/9811131].
%%CITATION = HEP-TH 9811131;%%
}

\lref\topo{N. Berkovits and C. Vafa, ``${\cal N}=4$ topological strings,''
Nucl. Phys. {\bf B433}, 123 (1995, [arXiv: hep-th/9407190].}

\lref\reviewhybrid{N. Berkovits, ``A New Description
of the Superstring,'' [arXiv: hep-th/9604123].}

%\GopakumarII
\lref\GopakumarII{
R.~Gopakumar and C.~Vafa,
``M-theory and topological strings. I,''
arXiv:hep-th/9809187.
%%CITATION = HEP-TH 9809187;%%
}
%\GopakumarJQ
\lref\GopakumarJQ{
R.~Gopakumar and C.~Vafa,
``M-theory and topological strings. II,''
arXiv:hep-th/9812127.
%%CITATION = HEP-TH 9812127;%%
}

 %\DijkgraafDH
\lref\DijkgraafDH{
R.~Dijkgraaf and C.~Vafa,
``A perturbative window into non-perturbative physics,''
arXiv:hep-th/0208048.
%%CITATION = HEP-TH 0208048;%%
}

\lref\cdeformc{
R. ~Dijkgraaf, M.T. ~Grisaru,
H. ~Ooguri, C. ~Vafa, and D. ~Zanon,
``Planar gravitational corrections for supersymmetric
gauge theories,''
%%CITATION = HEP-TH 0310061
arXiv:hep-th/0310061.}

%\cdeformb
\lref\cdeformb{
H.~Ooguri and C.~Vafa,
``Gravity induced $C$-deformation,''
arXiv:hep-th/0303063.
%%CITATION = HEP-TH 0303063;%%
}

%\cdeforma
\lref\cdeforma{
H.~Ooguri and C.~Vafa,
``The $C$-deformation of gluino and non-planar diagrams,''
arXiv:hep-th/0302109.
%%CITATION = HEP-TH 0302109;%%
}

%\OoguriBV
\lref\OoguriBV{
H.~Ooguri and C.~Vafa,
``Knot invariants and topological strings,''
Nucl.\ Phys.\ B {\bf 577}, 419 (2000)
[arXiv:hep-th/9912123].
%%CITATION = HEP-TH 9912123;%%
}
%\MarinoRE
\lref\MarinoRE{
M.~Marino and C.~Vafa,
``Framed knots at large $N$,''
arXiv:hep-th/0108064.
%%CITATION = HEP-TH 0108064;%%
}
%\LabastidaYW
\lref\LabastidaYW{
J.~M.~Labastida, M.~Marino and C.~Vafa,
``Knots, links and branes at large $N$,''
JHEP {\bf 0011}, 007 (2000)
[arXiv:hep-th/0010102].
%%CITATION = HEP-TH 0010102;%%
}
%\WittenYC
\lref\WittenYC{
E.~Witten,
``Phases of ${\cal N} = 2$ theories in two dimensions,''
Nucl.\ Phys.\ B {\bf 403}, 159 (1993)
[arXiv:hep-th/9301042].
%%CITATION = HEP-TH 9301042;%%
}
\newsec {Introduction}
There is by now a large class of examples in string theory
that realizes the idea of `t Hooft of large $N$ dualities
for gauge theories.  Most of the arguments for the existence
of such dualities derive from the target space perspective:
the back-reaction on the gravity modes by the D-branes.
However, the original motivation of `t Hooft was a statement visible
at the level of the worldsheet, namely he conjectured that somehow
the holes in the large $N$ expansions of Feynman diagrams
 close up and lead to a closed
string expansion.  Thus these dualities are expected to be visible genus by
genus
in the worldsheet.
  Understanding the large $N$ dualities from this viewpoint is crucial
because it also will teach us how the large $N$ dualities, unlike
U-dualities, are
derivable from perturbative considerations of closed string theory.

A simple example of large $N$ duality was proposed in \GopakumarKI\
which relates large $N$ Chern-Simons theory on $S^3$,
which is equivalent to open topological strings \WittenFB , with topological
closed strings on the resolved conifold, where
the size of the blown up ${\bf P}^1$ is given by the
`t Hooft parameter.  This duality has been derived
from a worldsheet perspective in \OoguriGX :  Starting from the closed
string side and using the linear sigma model description of the
worldsheet theory \WittenYC , one discovers that in the limit
of small `t Hooft parameter, the worldsheet develops a new phase
(the Coulomb phase) which leads to the emergence of the open
string description.  The new phase of the closed
string worldsheet corresponds to the `filled holes'
of the open string worldsheet.

On the other hand, motivated from the meaning of topological
string computations as $F$ term computations in an associated superstring
\BershadskyCX, this topological string duality was embedded
in superstrings \VafaWI , and extended to a relatively large class
of superstring dualities (see e.g. \CachazoSG ), and led to
a link between ${\cal N}=1$ supersymmetric gauge theories and matrix models
\DijkgraafDH .  Even though the worldsheet derivation
of the topological string duality would lead, by a chain of arguments,
to the $F$ term dualities in superstring context, a direct worldsheet derivation
of these dualities was missing in the context of the superstring.

In this
paper we aim to fill this hole, at least at the level of $F$ terms.
A $d=4$ spacetime-supersymmetric description of the superstring
on Calabi-Yau threefolds is given by the hybrid formalism \refs{\hybrid,
\topo, \reviewhybrid}, which
is related to the RNS formalism by a field redefinition. We will
show that the computation of $F$ terms using the hybrid formalism
is equivalent to the computation of $F$ terms using a ten-dimensional
topological string with $\hat c=5$.
We will then use the $\hat c=5$ topological string to establish
the worldsheet equivalence of $F$ terms between open and closed sides.
In particular, we will find using the $\hat c=5$ description that
the D brane boundary state for the ten-dimensional
open superstring naturally emerges on the worldsheet of
the closed superstring dual.

The topological string method has
been used in motivating some of the results on superpotential
terms in gauge theories,
for example in \refs{\cdeforma, \cdeformb, \cdeformc},
which have then been verified by field theory methods.
This paper provides a precise justification of
these results from the string theory perspective.
While we establish the equivalence of closed and open strings
only at the level of $F$ terms, the setup we present should be
viewed as the first step in the derivation of the full duality

The organization of this paper is as follows. In section 2 we review
the worldsheet derivation of large $N$ topological string duality
\OoguriGX .  In section 3 we formulate topological strings directly in
ten dimensions, with $\hat c=5$, and show its equivalence to
the hybrid formalism \refs{\hybrid ,
\topo,\reviewhybrid}\
when evaluating $F$ terms for superstring compactifications.
In section 4 we use this $\hat c=5$ topological
formulation of the superstring to establish
the worldsheet equivalence of $F$ terms between open and closed sides.

\newsec {Review of Topological String Duality}

In this section, we will briefly review the worldsheet derivation
\OoguriGX\
of the duality between the A-type topological closed string on the
resolved conifold and the open topological string on the deformed
conifold with $N$ A-branes wrapping on the $S^3$ of the conifold.
The topological string coupling constants are the same
on both sides of the duality and denoted by $\lambda$.
The K\"ahler moduli $t$ of the resolved conifold (the ``size'' of the
${\bf P}^1$) in the closed string side is mapped to the number $N$
of the A-branes in the open string side by the relation,
\eqn\topduality{ t = i N \lambda.}
In this sense, this is an example of the 't Hooft duality.
This duality was conjectured in \GopakumarKI , and
various evidences for the duality have been found in
 \refs{ \OoguriBV, \LabastidaZP , \RamadeviGQ ,\LabastidaYW, \LabastidaTS,
 \MarinoRE}.

To derive the duality, we start with the closed string side
and expand string amplitudes in powers of $t$. What is expected
to emerge from the duality is a sum over open
string worldsheets with each boundary weighted by the factor
of $N\lambda=-it$. The target space becomes singular in the limit
$t \rightarrow 0$,
and the worldsheet in the limit is
best described by using the linear sigma model \WittenYC .
For the resolved conifold, the linear sigma model consists of
four chiral multiplets, whose scalar fields are denoted by
$a_1, a_2$ and $b_1, b_2$, and one vector multiplet, whose
scalar field is denoted by $\sigma$. The chiral multiplet fields
$a_1, a_2$ carry charge $+e$ with respect to the gauge field $A$
in the vector multiplet, and $b_1, b_2$ carry $-e$. After integrating
out the auxiliary fields, the potential $U$
for the bosonic fields are given as
\eqn\potential{
U
= |\sigma|^2\left( |a_1|^2 + |a_2|^2 + |b_1|^2 + |b_2|^2 \right)
 + e^2\left( |a_1|^2 + |a_2|^2 - |b_1|^2 - |b_2|^2 \right).}

According to the duality relation \topduality ,
the K\"ahler moduli is pure imaginary.
In this case, $t$ appears as the theta term for the gauge field
$\sim t \int dA$. If we introduce a twisted chiral superfield
$\Sigma$ defined from the vector superfield $V$ as
$\Sigma = \bar D_+ D_- V = \sigma + \cdots$, the theta term
can be also written as as an $F$ term with the superpotential
\eqn\linearpotential{ W = t \Sigma.}
We will find this description in terms of $\Sigma$
to be useful in the following discussion.

When $t \neq 0$, the linear sigma model flows  in the infrared limit
to the non-linear sigma model for the conifold. The theta term
lifts the Coulomb branch and constrains $\sigma = 0$. Since
$e\rightarrow \infty$ in the infrared limit,
the chiral multiplet fields should obey $|a_1|^2 + |a_2|^2
= |b_1|^2 + |b_2|^2$ modulo the gauge symmetry, $(a_{1,2}, b_{1,2})
\rightarrow (e^{i\theta} a_{1,2}, e^{-i\theta} b_{1,2})$.
We recognize this quotient is  the conifold
geometry.\foot{The gauge invariant combinations, $z_{ij} = a_i b_j$,
obey the relation $z_{11} z_{22} - z_{12} z_{21} = 0$ defining the
conifold geometry.
For a given set of $z_{ij}$, the original fields $a_i$ and $b_i$
are determined modulo $(a_{1,2},b_{1,2}) \rightarrow
(e^\rho a_{1,2}, e^{-\rho} b_{1,2})$, which is taken into
account by the gauge symmetry and the constraint
$|a_1|^2 + |a_2|^2 = |b_1|^2 + |b_2|^2$.}
In this limit, $\sigma$ is identified with
the chiral primary field associated to the element of $H^{1,1}$ dual
to the ${\bf P}^1$.

When we expand around $t=0$, however, we need to take into account
a new flat direction where $\sigma$ can be non-zero. Due to the
potential \potential , the chiral multiplet fields are now constrained
to vanish, $a_{1,2}=b_{1,2}=0$. We call this flat direction as the
$C$ branch. In comparison, the branch where $\sigma=0$ is called
the $H$ branch. When we quantize the linear sigma-model, we need to
integrate over both $C$ and $H$ branches. It is useful to think that
the worldsheet is divided into $C$ and $H$ domains, where the
fields take values in the $C$ and $H$ branches respectively. Performing
the functional integral involves summing over all possible configurations
of these two branches.

We expect that quantization of the $H$ branch still leads to the
sigma-model on the conifold away from the conifold point. How to
remove the conifold point would depend on how we divide the
integral over $\sigma$ between the two branches.
On the other hand, the $C$ branch is non-geometric
since $a_{1,2}, b_{1,2}$, which are coordinates for the conifold,
become massive. We regard $C$ domains as holes on the worldsheet
and claim that this is how open strings emerge from the closed
string theory.
For this interpretation to work, we need that:

(1) Every $C$ domain has the topology of the disk.

~~~~Contributions from all other topologies should vanish
in string amplitudes.

(2) Each disk in the $C$ branch contributes the factor of $-it
=N\lambda$.

\noindent
It was shown in \OoguriGX\ that both of these statements are
true.

To show (1), it was noted that each $C$ domain contributes to a
topological string amplitude as
\eqn\cdomain{ \oint d\sigma_0 {\partial \over \partial \sigma_0}
{\cal F}^{(C)}(\sigma_0),}
where ${\cal F}^{(C)}(\sigma)$ is the partition function for the
$C$ domain with the boundary condition $\sigma = \sigma_0$.
The action of $\oint d\sigma_0 \partial/\partial \sigma_0$ is
due to a Jacobian factor that is needed to trade a part of the
functional integral into an integral over configurations of the
$C$ domain. By the topological BRST symmetry, ${\cal F}^{(C)}(\sigma_0)$
is holomorphic in $\sigma_0$. This means that the contribution
\cdomain\ would vanish if ${\cal F}^{(C)}$ is a single-valued function
of $\sigma_0$. This is the case when the $C$ domain has a handle
or more than one boundaries. The only exception is the case
when the $C$ domain has the topology of the disk. The string
amplitude on the disk is not well-defined unless we have some
punctures, and ${\cal F}^{(C)}(\sigma_0)$ can have a monodromy
around $\sigma_0=0$, which can be picked up by the integral
in \cdomain .

To evaluate \cdomain , we note that the $C$ domain has a description
as a Landau-Ginzburg model with the superpotential $W$ being
given by \linearpotential . The disk amplitude is then given
by an integral of $\exp(-W)$. The only subtlety is the measure factor
of $\sigma^{-2}$ which arises from the integral over $a_{1,2},
b_{1,2}$, which are massive in this domain. Taking this into account,
we find,
$$ {\cal F}^{(C)}(\sigma_0)  = \int^{\sigma_0} {d \sigma \over \sigma^2}
\exp(- t\sigma). $$
This show that the disk amplitude is indeed multivalued around
$\sigma_0=0$ as ${\cal F}^{(C)}(\sigma_0) \sim -\sigma_0^{-1}
-t\log\sigma_0 +\cdots$. Therefore the contribution of the
$C$ domain of the disk topology is given by
$$ \oint d\sigma_0 {\partial \over \partial \sigma_0}
{\cal F}^{(C)}(\sigma_0)
= \oint {d\sigma_0 \over \sigma_0^2} \exp(-t \sigma)
\sim -it=N\lambda.$$
This shows that (1) and (2) are indeed true for the
closed string theory.

We have found that the closed string amplitude, when expanded
in powers of $t$, can be expressed as a sum over holes on
the worldsheet with the power of $t$ keeping track of the
number of holes. Namely the closed string theory is indeed
equivalent to an open string theory with some boundary
condition. Is the boundary condition exactly what we expect
from the large $N$ duality? Since the worldsheet variables
$a_{1,2}, b_{1,2}$ become massive in the $C$ domain,
 near the interface of the $C$ and $H$ domains, they
stay near the tip of the conifold.
 Their precise behavior depend on how we
divide the $\sigma$ integral between the two branches.
On the other hand, the A brane for the open string
is supposed to wrap on the $S^3$ of the deformed conifold.
Its size is undetermined since changing the radius is a
BRST trivial deformation. When the radius is small, the $S^3$ is
near the tip of the conifold. Therefore, modulo the ambiguities
that exist in both sides of the duality, the boundary of the
$C$ domain correctly reproduces the A brane boundary condition
in the open string dual.

\newsec {Equivalence of $\hat c=5$ and Hybrid Computation of $F$ Terms}

In this section we introduce the concept of topological strings
in ten dimensions  with $\hat c=5$, generalizing
the topological strings often used in the context of Calabi-Yau
threefolds, and establish
its direct equivalence
to the hybrid formalism for certain
 $F$ term computations in type II superstrings.

In the first subsection, we will show that states in the $G^+$
cohomology in the $\hat c=5$
topological string include
supersymmetry
multiplets containing massless compactification moduli as well
as the multiplet containing the self-dual graviphoton field strength.
In the second subsection,
we will give a $\hat c=5$ topological prescription for computing
tree and loop scattering
amplitudes involving these states which will
contribute only to $F$ terms in the low-energy effective
action. And in the third
subsection, we will show how to describe these states using
the hybrid formalism and will prove that the hybrid prescription for
their
scattering amplitudes agrees with the $\hat c=5$ topological
string prescription.

\subsec{Chiral states using the $\hat c=5$ description}

The worldsheet fields in the $\hat c=5$ formalism include the $d=4$
variable $x^m$ for $m=0$ to 3, the left-moving chiral superspace variables
$\t^\a$ and its conjugate momentum $p_\a$ for
$\a=1$ to 2, and
an ${\cal N}=2$ $\hat c=3$ superconformal field theory for the internal
compactification manifold.
Unlike the superstring in the hybrid formalism,
the $\hat c=5$ formalism
does not involve dotted superspace variables
$\t^{*\ad}$ or its conjugate momenta $p^*_\ad$, and also does not contain
the chiral boson $\rho$.
For the type II superstring, the $\hat c=5$ formalism also includes
the right-moving fermionic variables $\tbar^\a$ and its conjugate momenta
$\bar p_\a$, but does not involve $\tbar^{*\ad}$ or $\bar p^{*\ad}$.
(We will reserve barred notation throughout
this paper to denote
right-moving variables, and will use the $*$ superscript to denote
dotted spinor variables.)
For the formalism to be Hermitian, one therefore needs to Wick-rotate
to either
signature $(4,0)$ or $(2,2)$ so that $\t^\a$ is real. Although
the reality conditions for spacetime fields in these signatures
are not the standard ones, it
is straightforward to Wick-rotate back to the standard Minkowski reality
conditions after
computing scattering amplitudes and determining
the corresponding $F$ terms in the
effective action.

In the ${\cal N}=2$ $\hat c=5$ formalism, the worldsheet action is
$$S=\int d^2 z (p^\a \bar\p \t_\a +\bar p^\a\p\bar\t_\a
+\half\e_{\a\b}\p x^{\a \dot +}\bar\p x^{\b \dot -}) + S_{CY}$$
and the
left and right-moving twisted ${\cal N}=2$ generators are
\eqn\chatgen{
\eqalign{&T= p_\a \p\t^\a + \half\e_{\a\b}\p x^{\a \dot +}\p x^{\b \dot -}
+ T_{CY},\cr
& G^+ = \t_\a \p x^{\a \dot +} + G^+_{CY},~~~
 G^- = p_\a \p x^{\a \dot -} + G^-_{CY}, \cr
& J = \t^\a p_\a + J_{CY}. \cr
&\bar T=
\bar p_\a \bar
\p\bar \t^\a + \half\e_{\a\b}\bar \p x^{\a \dot +}\bar \p x^{\b \dot -}
+\bar  T_{CY},\cr
&\bar G^+ = \bar\t_\a \bar\p x^{\a \dot +} +\bar G^+_{CY},~~~
\bar G^- =\bar p_\a \bar\p x^{\a \dot -} +\bar G^-_{CY},\cr
&\bar J = \bar\t^\a\bar p_\a +\bar J_{CY},}}
where $x^{\a\ad}= x^m \s_m^{\a\ad}$ and $\ad=(\dot +,\dot -)$,
$S_{CY}$ and
$\{~T_{CY}, G^+_{CY},G^-_{CY},J_{CY}~\}$ are the worldsheet action
and twisted ${\cal N}=2$ $\hat c=3$ generators
for the internal compactification manifold,
and $G^+$ and $G^-$ carry conformal
weight $+1$ and $+2$ respectively.
In the traditional description of the topological string,
one treats
$(x^{\a \dot +} ,\t^\a,\bar\t^\a)$ as holomorphic coordinates
on ${\bf C}^2={\bf R}^4$ and their superpartners
and $(x^{\a \dot -}, p^\a, \bar p^\a)$ as anti-holomorphic coordinates
and their partners.
The four-dimensional part of the twisted ${\cal N}=2$ theory is
then the topological B model whose target space is ${\bf C}^2$.
Note that the ${\cal N}=2$ generators of \chatgen\ only preserve
a $U(1)\times SU(2)$ (or $GL(1)\times SL(2)$)
subgroup of $SO(4)$ (or $SO(2,2)$) Lorentz invariance in
the signature $(4,0)$ (or $(2,2)$).
For simplicity, we will usually
restrict our attention to the
left-moving sector.

Since $\oint G^+= \oint( \t_\a \p x^{\a \dot +} + G^+_{CY})$
plays the role of a BRST operator in the topological ${\cal N}=2$
string, it is natural to compute its cohomology.
Since $\t_\a \p x^{\a\dot +}$ and $G^+_{CY}$ involve different worldsheet
fields, states $V$ in the cohomology of $\oint G^+$ can be written as
$V= \sum_i \Phi^i \sigma_i$
where $\Phi^i$ is constructed from the four-dimensional fields
$\{ x^m,\t^\a,p_\a \}$ and
is in the cohomology of $\oint \t_\a \p x^{\a\dot +}$,
and $\sigma_i$ is constructed from compactification-dependent fields
and is in the cohomology of $\oint G^+_{CY}$. Using the standard quartet
argument, states in the cohomology of $\oint\t_\a \p x^{\a\dot +}$
can depend only on the zero modes of $\t^\a$ and $x^{\a \dot +}$.
So the most general state in the
cohomology of $\oint G^+$ is
\eqn\chirals{V= \sum_i \Phi^i (x^{\a \pd},\t^\b,\bar\t^\g) ~\sigma_i}
where $\sigma^i$ is
in the cohomology of $\oint G_{CY}^+$.
Such states will be called ``chiral'' states.

In this paper, we shall only consider chiral states where $\sigma_i$
contains either $+1$ or zero $U(1)$ charge with respect to the
left and right-moving internal $J_{CY}$.
($\sigma_i$ carrying zero internal $U(1)$ charge
correspond to the identity operator.)
Note that the $U(1)$ charge in the $d=4$
sector is unconstrained in the chiral states considered here.

For the Type IIA (or Type IIB)
superstring, chiral states carrying $+1$ left and
right-moving $U(1)$ charge
in the internal
sector correspond to massless multiplets associated with
K\"ahler (or complex)
moduli of the Calabi-Yau space. The associated chiral
moduli vertex operator is
\eqn\moduli{V= \sum_i \Phi^i(x^{\a\pd},\t,\bar\t) ~\sigma_i }
where $\sigma_i$ is a chiral primary of (left,right)-moving
charge $(+1,+1)$ associated with
the internal ${\cal N}=2$ $\hat c=3$ superconformal field theory.
The $\t=\tbar=0$ component of $\Phi^i$ is the chiral modulus field
and the
$\t=\tbar=0$ component of $D_\a \bar D_\b\Phi^i$ is the self-dual
Ramond-Ramond (R-R) flux associated with this modulus.

For both the Type IIA and IIB superstring,
chiral states carrying zero $U(1)$ charge in the
internal sector correspond to
a multiplet containing the self-dual graviphoton.
The associated self-dual graviphoton vertex operator is
\eqn\grav{V= R(x^{\a\pd},\t,\tbar) }
where the self-dual graviphoton field strength $F_{\a\b}$ is
the $\t=\bar\t=0$ component of
$\p_{\a \pd}\p_{\b \pd} R$ and the self-dual Riemann tensor $R_{\a\g\b\d}$
is the
$\t=\bar\t=0$ component of
$\p_{\a \pd}\p_{\b \pd}D_\g \bar D_\d R$.

Although the chiral states of \moduli\ and \grav\ do not have fixed
charge with respect to the $U(1)$ charges $\int dz J$ and $\int d\bar z
\bar J$ of \chatgen, they can be defined to have fixed charge with respect
to
\eqn\newcharge{\int dz (J+K) +\int d\bar z (\bar J+\bar K)}
where $K= \half\epsilon_{\a\b} x^{\a\pd} \p x^{\b\md}- \t^\a p_\a$ and
$\bar K=
\half\epsilon_{\a\b} x^{\a\pd} \bar\p x^{\b\md} -\bar\t^\a\bar p_\a$.
Note that $\int dz K +\int d\bar z \bar K$ is a conserved charge which
commutes with the ${\cal N}=2$ generators of \chatgen. When \moduli\ is
independent of $x^{\a\pd}$ and
\grav\ is quadratic in $x^{\a \pd}$ (i.e. when
$F_{\a\b}$ and $R_{\a\b\g\d}$ are constants),
these chiral states all have charge $+2$ with respect to \newcharge.

\subsec{Scattering amplitudes using the $\hat c=5$ formalism}

To compute scattering amplitudes of chiral states using the $\hat c=5$
formalism, we shall use the topological ${\cal N}=2$ prescription where
$\oint G^+$ is treated as the BRST charge and $G^-$ is treated as the
$b$ ghost. For $M$-point $g$-loop Type II scattering amplitudes, the
${\cal N}=2$ topological prescription is
\eqn\topamp{
A_{g,M}=
\left\langle \big| \prod_{j=1}^{3g-3+M}\int dm^j\int \mu_j G^-\big|^2
\quad \prod_{r=1}^M V_r(z_r) \right\rangle }
where $\mu_j$ denotes the $(3g-3+M)$ Beltrami differentials associated with
the worldsheet moduli $m_j$,
and $\big|~\big|^2$ signifies the product of left and right-moving
terms. Since $\hat c=5$, this amplitude vanishes by charge conservation
unless
\eqn\charger{5(1-g) = \sum_{r=1}^M J_r - (3g-3+M),}
where $J_r$ is the $U(1)$ charge of $V_r$.
So the sum of the $U(1)$ charges of the vertex operators
must be equal to $(2-2g+M)$ both in the left and right-moving sectors.

The $M$-point $g$-loop amplitudes considered here will involve
$(M-2g)$ chiral moduli described by the vertex operators of \moduli\ and
$2g$ self-dual graviphoton vertex operators described by the vertex operators
of \grav. With this choice, the charge conservation equation of
\charger\ implies that $+2$ left and right-moving $U(1)$ charge must
come from the $d=4$ sector of the formalism. As will be seen below,
this $d=4$ $U(1)$ charge comes from the zero modes of $\t^\a$ and $\tbar^{\a}$.
Although it might be interesting to consider more general
scattering amplitudes in the $\hat c=5$ formalism,
it is not clear if more general $\hat c=5$ scattering amplitudes will be
$d=4$ super-Poincar\'e invariant like the amplitudes considered here.

In computing these special scattering amplitudes, it will be convenient
to choose $2g$ of the $(3g-3+M)$ Beltrami differentials
to be associated with the
locations of the graviphoton vertex operators. So the formula of
\topamp\ becomes
\eqn\topamptwo{
A_{g,M}=
\left\langle \big| \prod_{j=1}^{g-3+M}\int dm^j\int \mu_j G_{CY}^-\big|^2
\quad \prod_{r=1}^{M-2g}\Phi^{i_r}_r \sigma_{i_r}(z_r) \prod_{s=1}^{2g}
\int d^2 z_s W_s(z_s) \right\rangle }
where
\eqn\intgrav{
\eqalign{W_s =&
\oint G^-\oint \bar G^- R(x,\t,\tbar)
\cr =&
(p^\a \p_{\a\pd}+\p x_{\a\pd}{\p\over{\p\t_\a}})
(\bar p^\b \p_{\b\pd}+\bar\p x_{\b\pd}{\p\over{\p\tbar_\b}}) R(x,\t,
\tbar),}}
and $\oint G^-\oint \bar G^- R$ signifies the single pole of $G^-$ and
$\bar G^-$ with $R$.
It will be useful to note that since
$(3-3g)$ $U(1)$ charge is needed from the internal sector,
only the $G_{CY}^-$ term in $G^-$ contributes in
$\int \mu_j G^-.$

In order that the $\mu_j G^-$ integrals in \topamp\ reproduce the correct
Faddeev-Popov measure for integration over worldsheet metrics, it is
usually required that the vertex operators $V_r$
have no double (or higher-order)
poles with $G^-$. This condition guarantees that $\oint G^- V$
has no singularities
with $G^-$ which, together with $\oint G^+ V=0$, implies that
$V$ is an ${\cal N}=2$ chiral primary.
For chiral states of the two types considered here, this would imply that
\eqn\impose{{\p\over{\p\t_\a}} \p_{\a \pd} \Phi^i=0\quad {\rm and}
\quad
{\p\over{\p\t_\a}} \p_{\a \pd} R=0.}
However, for the amplitudes considered here,
these conditions are unnecessary since
only the $G^-_{CY}$ term contributes in $\int \mu_j G^-$.
So there is no problem with reproducing the
Faddeev-Popov measure if the vertex operators
in \topamptwo\ have singularities with the $d=4$ part of $G^-$, and there
is therefore no need to impose \impose\ for consistency
of these scattering amplitudes.

Furthermore, the fact that only $G^-_{CY}$ contributes to $\int \mu_j G^-$
implies that the amplitude is spacetime supersymmetric. To show this, define
the spacetime supersymmetry generators in the $\hat c=5$ formalism as
\eqn\chatsusy{q_\a = \oint p_\a, \quad q^*_\ad = \oint \t^\a \p x_{\a\ad},}
which anticommute to the usual
supersymmetry
algebra
$$\{q_\a,q_\b\}=0,\quad
\{q^*_\ad,q^*_\bd\}=0,\quad
\{q_\a,q^*_\bd\}=\oint \p x_{\a\bd}.$$
Note that these supersymmetries preserve the $\oint G^+$ cohomology
when acting on states that carry no $P^{\a \pd}$ momentum
since $\{q^*_\ad,\oint G^+\}=0$ and
$\{q_\a, \oint G^+\}=\int \p x^{\a \pd}.$
Finally, note that $\{q_\a, G^-\}= \{q^*_\ad, G_{CY}^-\}= 0$ and
$\{q^*_\ad, G_{4d}^-\} = \d_\ad^\pd~ T_{4d}$ where $G^-_{4d}$ and $T_{4d}$
are the
four-dimensional contributions to $G^-$ and $T$.
Since $G_{4d}^-$ appears only in the integrated graviphoton vertex operator of
\intgrav, the anticommutator
$\{q^*_\ad, G_{4d}^-\}
= \d_\ad^\pd~ T_{4d}$ can be ignored since it only shifts
the graviphoton vertex operator by a surface term.

To obtain the supersymmetric $F$ term associated with the amplitude of
\topamptwo, integrate over the zero modes of $(x^m,\t^\a,\tbar^\a)$
and use the graviphoton vertex operator of \intgrav\ to absorb the zero
modes of $p^\a$. In terms of the self-dual graviphoton superfield
$F_{\a\b}=\p_{\a\pd}\p_{\b\pd} R$, one finds
\eqn\topampthree{
\eqalign{
A_{g,M}=&\int d^4 x \int d^2\t \int d^2\tbar
\prod_{r=1}^{M-2g}\Phi^{i_r}_r(x,\t,\tbar)
\prod_{s=1}^{2g} F_{s~\a\b}(x,\t,\tbar)  \cr
&\times 
\left\langle \big| \prod_{j=1}^{g-3+M}\int dm^j\int \mu_j G_{CY}^-\big|^2
\quad \prod_{r=1}^{M-2g} \sigma_{i_r}(z_r) \right\rangle_{CY} }}
where
$\left\langle ~~\right\rangle_{CY}$ denotes a
functional integral over the
internal compactification-dependent fields and
the $2g$ $\a$ indices and $2g$ $\b$ indices in $\prod_{s=1}^{2g}
F_{s~\a\b}$
are contracted with each other in all
possible combinations.
So the $F$ term associated with this scattering amplitude is
\eqn\Fchat{S= f_{i_1 ... i_{M-2g}}~ \int d^4 x \int d^2\t \int d^2\tbar
\left(F_{\a\b}(x,\t,\tbar) F^{\a\b}(x,\t,\tbar)\right)^g
\prod_{r=1}^{M-2g}\Phi^{i_r}(x,\t,\tbar) }
where the coefficient $f_{i_1 ... i_{M-2g}} $ is
defined by the ${\cal N}=2$ $\hat c=3$ topological
amplitude
$$f_{i_1 ... i_{M-2g}}  =
\left\langle \big| \prod_{j=1}^{g-3+M}\int dm^j\int \mu_j G_{CY}^-\big|^2
\quad \prod_{r=1}^{M-2g} \sigma_{i_r}(z_r) \right\rangle_{CY} .$$
If we denote the K\"ahler (complex) moduli by $t_i$ and denote
the topological string amplitude at genus $g$ by $F_g(t_i)$, then
$$f_{i_1 ... i_{M-2g}}=\partial_{i_1}...\partial_{i_{M-2g}}F_g(t_i).$$

\subsec{Hybrid description of chiral states}

It will be shown here that the scattering amplitudes of chiral moduli states
and self-dual graviphoton states computed in \topampthree\ using the
$\hat c=5$ formalism agree with those computed using the hybrid formalism.
Note that hybrid scattering amplitudes involving only self-dual graviphoton
states were computed previously in
\topo.
As discussed
in \refs{\hybrid ,\topo,\reviewhybrid},
the hybrid formalism is
related to the RNS formalism by a field redefinition.
In the hybrid formalism, physical superstring states
are described by chiral primary fields of $+1$ $U(1)$ charge
with respect to the twisted ${\cal N}=2$ $\hat c=2$
generators
\eqn\GSf{
\eqalign{&T=\half\dzm x^m \dzm x_m +
p_\a\dzm \t^\a +  p^*_\ad \dzm\tb^\ad +\half\dzm\rho\dzm\rho +\half\p^2\rho
+ T_{CY}, \cr
&G^+=e^{-\rho} ( d^*)^2 + G^+_{CY}, ~~
G^-=e^{\rho} (d)^2 +G^-_{CY} , \cr
&J=\dzm\rho + J_{CY}, }}
where
$$d_\a=p_\a+i\tba\dzm x_{\a\ad}-(\tb)^2\dzm\t_\a,\quad
d^*_\ad= p^*_\ad,$$
and $\{~T_{CY},G^+_{CY},G^-_{CY},J_{CY}~\}$ are the same twisted
${\cal N}=2$ $\hat c=3$ generators
as before. Note that $\rho$ is a negative-energy
chiral boson satisfying the OPE
$$\rho(y)\rho(z)\sim -\log(y-z)$$
and $d_\a$ and $d^*_\ad$ are defined
such that they
anticommute with the supersymmetry generators
$$q_\a =\oint p_\a, \quad q^*_\ad = \oint( p^*_\ad -i\t^\a \p x_{\a\ad})$$
and satisfy the OPE's
\eqn\dope{d_\a(y) d^*_\ad(z)\sim {1\over y-z}
 (\p x_{\a\ad} +i\t^*_\ad\p\t_\a).}

To compare scattering amplitudes using the hybrid formalism
with those of \topampthree, one first
needs the hybrid version of the
vertex operators for the chiral moduli and graviphoton multiplets.
The superstring states corresponding
to compactification moduli multiplets are described in the hybrid
formalism by the vertex operators
\eqn\hymoduli{V= \sum^i \Phi^i(x,\t,\bar\t)\sigma_i, }
where $\sigma_i$ is the same compactification-dependent field as in
the $\hat c=5$ description and carries
$+1$ left and right moving $U(1)$ charge.
One can easily check that $V$ is chiral ($i.e.$
 is annihilated by $\oint G^+$ and $\oint \bar G^+$) if
$D^{*\ad}\Phi^i=\bar D^{*\ad}\Phi^i=0$ and is a chiral primary ($i.e.$
has no double poles with $G^-$) if $D_\a D^\a \Phi^i=\bar D_\a 
\bar D^\a \Phi^i=0$.

Because of the
additional condition $D^\a D_\a\Phi^i=\bar D^\a \bar D_\a \Phi^i=0$, 
the $\hat c=5$ vertex operator
$V=\sum_i\Phi^i~\sigma_i$ is not necessarily a chiral primary vertex operator
in the hybrid formalism. However, as will be seen later in this subsection,
the condition $D^\a D_\a\Phi^i=\bar D^\a \bar D_\a \Phi^i=0$ 
will not be necessary for consistency of
hybrid scattering amplitudes involving only chiral states.
This is because, just as in the $\hat c=5$ formalism,
only the $G^-_{CY}$
term will contribute in $G^-$ for these scattering amplitudes in the
hybrid formalism. So there is no problem if the vertex operators
have singularities with the four-dimensional $d^2 e^\rho$ term in $G^-$.
This implies that one can prove equivalence of scattering amplitudes
even for chiral states such as
$V=(\t-\tbar)^\a (\t-\tbar)_\a ~ \sigma$ which are not ${\cal N}=2$
primary fields in the hybrid formalism and therefore do not correspond
to on-shell superstring states.
This vertex operator $V$, which corresponds to a supersymmetric
combination of the R-R and NS-NS fluxes associated to the moduli $\sigma$,
will play an important role in the next section.

The superstring state corresponding to the self-dual graviphoton multiplet
will be described in the hybrid formalism by the vertex operator
\eqn\hygrav{
V=  e^{-\rho} p^{*\pd} e^{-\bar\rho} \bar p^{*\pd} R(x,\t,\tbar).}
This vertex operator is chiral if $D^*_\ad R = \bar D^*_\ad R=0$
and is primary if $D_\a\p^{\a \pd} R = \bar D_\a \p^{\a\pd} R=0$.
Although this vertex operator carries zero $U(1)$ charge in the internal
sector, it carries $+1$ left and right-moving $U(1)$ charge in the
four-dimensional sector because of its $\rho$ dependence. 
Using the OPE's of \dope,
one finds that
the integrated form of the graviphoton vertex operator is
\eqn\hyintgrav{
\eqalign{&\oint G^-\oint \bar G^- V \cr
&=
\int d^2 z\left( d_\a \p^{\a\pd}+(\p x^{\a \pd} +\t^{*\pd}\p\t^\a)D_\a
\right)
\left( \bar d_\b \p^{\b\pd}+(\bar\p x^{\b\pd} +\tbar^{*\pd}\p\tbar^\b)
\bar D_\b\right) R.}}
So if one sets $\t^*_\ad=\tbar^*_\ad=0$, this expression coincides
with the $\hat c=5$ expression of \intgrav.

To compute scattering amplitudes in the hybrid formalism, one first
extends the $\hat c=2$ ${\cal N}=2$ generators
of \GSf\ to a set of ${\cal N}=4$ generators
$$\{~T,G^+,\tilde G^+, G^-,\tilde G^-, J^{++},J, J^{--}~\}$$
by defining
$$J^{++}\equiv \exp \left(\int^z J\right),
\quad\quad J^{--}\equiv \exp \left(-\int^z J\right)$$
to form an $SU(2)$ set of generators together with $J$,
and by defining
$$\tilde G^- \equiv \left[\oint J^{--}, G^+\right] ,\quad
\tilde G^+ \equiv \left[\oint J^{++}, G^-\right],$$
to transform together with $G^+$ and $G^-$
as two doublets under this $SU(2)$. As discussed in
\topo, the $M$-point $g$-loop amplitude is defined by the formula
\eqn\ampfour{
\eqalign{& A_{M,g}(u_1,u_2,\bar u_1,\bar u_2) \cr
&=\prod_{i=1}^g \int d^2 v_i
\left\langle \big|\prod_{i=1}^{g-1} \widehat{{\tilde G}^+}(v_i) J(v_g)
\prod_{j=1}^{3g-3+M} dm^j
\int\mu_j \widehat{G^-} \big|^2
V_1 ... V_M \right\rangle,}}
where
$$\widehat {G^-} = u_1 G^- + u_2 \tilde G^-,\quad
\widehat{{\tilde G}^+} = u_1 \tilde G^+ + u_2 G^+,$$
and
$$\eqalign{
&A_{g,M}(u_1,u_2,\bar u_1,\bar u_2) \cr
&=
\sum_{P=2-2g-M}^{2g-2} \sum_{\bar P=2-2g-M}^{2g-2}
 (u_1)^{P+2g-2+M} (u_2)^{2g-2-P}
 (\bar u_1)^{P+2g-2+M} (\bar u_2)^{2g-2-\bar P} A_{g,M,P,\bar P}}$$
is a polynomial of degree $(4g-4+M,4g-4+M)$ in $(u,\bar u)$.
The different components $A_{g,M,P,\bar P}$ correspond to amplitudes
which violate (left,right)-moving $R$-charge by
$(P,\bar P)$. Note that $R$-charge in the hybrid formalism is equivalent
to picture in the RNS formalism.

For scattering amplitudes corresponding to $F$ terms with $(M-2g)$
chiral moduli and $2g$ graviphoton superfields, $R$-charge is violated
by $(g-1,g-1)$. This is because
chiral moduli superfields carry zero $R$-charge, self-dual
graviphoton superfields carry $(\half,\half)$ $R$-charge, and $F$ terms
carry $(-1,-1)$ $R$-charge from the $d^2\t d^2\tbar$ integration. So
we are interested in computing the component which violates $R$-charge
by $(P,\bar P)=(g-1,g-1)$.
To compute the $A_{g,M,g-1,g-1}$ component
of $A_{g,M}$ using
the formula of \ampfour, first note that all terms in this component
contain an equal number
of $\tilde G^-$ and $\tilde G^+$ operators.
To compare with the $\hat c=5$ prescription of \topamptwo,
it will be useful to first
turn all pairs of $(\tilde G^+,\tilde G^-)$ operators
into pairs of $(G^+,G^-)$ operators by performing the appropriate
contour deformations.

For example, suppose one has a pair of $\tilde G^+(y_1) \tilde G^-(y_2)$
operators at $y_1$ and $y_2$. First write $\tilde G^-=
[\oint G^+,J^{--}(y_2)]$ and deform the $\oint G^+$ contour off of
$J^{--}(y_2)$ until it hits the $J(v_g)$ operator, turning it into
$G^+(v_g)$. Secondly, write
$\tilde G^+(y_1)=[\int \tilde G^+,J(y_1)]$ and deform the $\int \tilde G^+$
contour off of
$J(y_1)$ until it hits the $J^{--}(y_2)$ operator, turning it into
$G^-(y_2)$. Finally, write
$G^+(v_g)=[\oint G^+,J(v_g)]$ and deform the $\oint G^+$
contour off of
$J(v_g)$ until it hits the $J(y_1)$ operator, turning it into
$G^+(y_1)$. So this procedure has turned
$\tilde G^+(y_1) \tilde G^-(y_2)$
into
$G^+(y_1) G^-(y_2)$.

In performing these contour deformations, we have ignored
possible surface terms on the moduli space of the worldsheet
coming from the commutator $[\oint G^+,\int \mu_j G^-]=
\int \mu_j T$,
where $\int \mu_j T$ produces a total derivative on the moduli space.
However, for the scattering amplitudes discussed here, one can
show that internal $U(1)$ charge conservation implies that these surface
terms do not contribute. As in the $\hat c=5$ computation, internal $U(1)$
conservation implies that the $d=4$ part of $G^-$ only contributes to the
scattering amplitude when it acts on the graviphoton vertex operator. Also,
one can argue by internal $U(1)$ conservation that only the $d=4$
part of $G^+$ contributes. So the only possibility of producing a surface
term comes from
$[\oint G^+_{4d},\int \mu_j G^-_{4d}]=\int \mu_j T_{4d}$
where the subscript $4d$ denotes
the four-dimensional contribution to these generators and
$\mu_j$ is associated
with the location of the graviphoton vertex operator. But this type of
surface term is harmless since it does not involve the $(3g-3)$
worldsheet moduli whose
boundary describes degeneration of the genus $g$ surface.

After replacing all $(\tilde G^+,\tilde G^-)$ pairs with
$(G^+,G^-)$ pairs and
choosing $2g$ of the Beltrami differentials to be associated
with the locations of the graviphoton vertex operators, one obtains
the formula
\eqn\ampfourtwo{
\eqalign{A_{M,g} = &
\prod_{i=1}^g \int d^2 v_i
\left\langle \big|\prod_{i=1}^{g-1} G^+(v_i) J(v_g)
\prod_{j=1}^{g-3+M}
\int dm^j \int\mu_j G^-\big|^2  \right. \cr
&~~~~~~~~~~~~~~~~\times \left.
\prod_{r=1}^{M-2g}\Phi^{i_r}_r \sigma_{i_r}(z_r) \prod_{s=1}^{2g}
\int d^2 z_s W_s(z_s) \right\rangle_H }}
where $W_s$ is defined in \hyintgrav\ and $\left\langle ~~\right\rangle_H$
denotes the functional integral using the hybrid formalism which
includes the $(\t^*_\ad,p^*_\ad)$ and $\rho$ fields.

To compare this formula with the $\hat c=5$ formula of \topamptwo,
insert the identity operator
$1= [\oint G^+, \t^*_\ad \t^{*\ad}e^\rho(w)]$ in \ampfourtwo\
and pull the $\oint G^+$ contour off of
$\t^*_\ad \t^{*\ad}e^{\rho}(w)$ until it hits $J(v_g)$ to give the formula
\eqn\ampfourthree{
\eqalign{A_{M,g} =&
\prod_{i=1}^g \int d^2 v_i
\left\langle \big|(\t^*_\ad \t^{*\ad} e^{\rho})(w)
\prod_{i=1}^g (p^{*\ad} p_\ad^* e^{-\rho})(v_i)
\prod_{j=1}^{g-3+M} \int dm^j
(\int\mu_j G_{CY}^-) \big|^2 \right. \cr
&~~~~~~~~~~~~\times \left.
\prod_{r=1}^{M-2g}\Phi^{i_r}_r \sigma_{i_r}(z_r) \prod_{s=1}^{2g}
\int d^2 z_s W_s(z_s) \right\rangle_H .}}
To derive \ampfourthree, we have used that $U(1)$ charge conservation
implies that only $G^-_{CY}$ contributes in the $\mu_j G^-$ terms and that
only $G^+_{4d}$ contributes to $G^+(v_i)$.

Finally, one needs to do the functional integral over the worldsheet
fields $(\t^*_\ad,p^*_\ad,\rho)$ which are present in the hybrid formalism but
not in the $\hat c=5$ formalism.
Since all $p^*_\ad$ variables in $G^+(v_i)$ must be
used to soak up the $2g$ zero modes of $p^*_\ad$, none of the
$\t^*_\ad$ variables in the vertex operators can contribute and
the
$\t^*_\ad \t^{*\ad}(w)$ soaks up the zero modes of $\t^*_\ad$. Because
the $\rho$ chiral boson has negative energy (like the $\phi$ chiral boson in
the RNS formalism which comes from fermionizing the $(\beta,\gamma)$
ghosts), it is subtle to define its functional integral. However,
for the amplitudes being considered here, the $\rho$ field always appears
together with the $(\t^*_\pd,p^{*\pd})$ fields in the combination
$\t^*_\pd e^\rho$ or $p^{*\pd} e^{-\rho}$. For this reason,
the functional integral over the $\rho$ chiral boson precisely cancels
the functional integral over the $(\t^*_\pd, p^{*\pd})$, even for
the zero modes. So after performing the functional integral over the
$(\t^*_\ad,p^*_\ad,\rho)$ fields, one obtains the amplitude
\eqn\ampfourfour{A_{M,g} =
\left\langle
\big| \prod_{j=1}^{g-3+M}\int dm^j
(\int\mu_j G_{CY}^-) \big|^2
\quad \prod_{r=1}^{M-2g}\Phi^{i_r}_r \sigma_{i_r}(z_r) \prod_{s=1}^{2g}
\int d^2 z_s W_s(z_s) \right\rangle ,}
which agrees with the $\hat c=5$ formula of \topamptwo.

\newsec{Large N Duality in Superstring}

It was pointed out in \VafaWI\ that the duality between the
open and closed topological string theories can be uplifted
to the type IIA superstring on the conifold times ${\bf R}^4$
with $N$ D5 branes wrapping on the ${\bf P}^1$ of the conifold and
extended in the ${\bf R}^4$ direction to another compactification
with $N$ units of R-R flux and without D branes. As far as
the $F$ terms are concerned, this superstring
duality is inferred
from the topological string duality combined with the relation
between the superpotential terms and the topological string
amplitudes \refs{\BershadskyCX , \OoguriBV}. This duality
is supposed to hold beyond the superpotential computation,
along the line of construction described in the closely related
papers  \refs{\KlebanovHB, \MaldacenaYY}. A derivation of the full
duality would require controlling back-reactions of the R-R fluxes
to the metric and understanding worldsheet dynamics in such
a background, and it would be tantamount to proving the AdS/CFT
correspondence. In this section, we will make the first step
in this direction by giving a direct worldsheet derivation
of the duality restricted to the superpotential computation,
where the back-reaction to the metric can be ignored as being
a BRST trivial deformation of the background.

As we saw in the last section,
the $\hat c=5$ formalism allows us to compute superpotential
terms as topological string amplitudes. In this formalism,
in addition to the $\hat c=3$ model discussed in section 2, we
have four bosons $x_{\alpha\dot\alpha}$ and four pairs of fermions
$(p_\alpha, \theta^\alpha)$ and $(\bar p_\alpha, \bar\theta^\alpha)$.
In the $\hat c=3$ model on the Calabi-Yau space, basic observables
are associated to cohomology elements of the Calabi-Yau space.
For example, for $\omega \in H^{1,1}$, we have $\sigma = \omega_{i\bar j}
\psi_L^i \psi_R^{\bar j}$. In the $\hat c=5$ formalism, it
can be multiplied
by any function of $\theta, \bar\theta$ as
$\Phi(\theta,\bar\theta)~ \sigma$, giving rise
to a vertex operator for the ${\cal N}=2$ vector multiplet in four
dimensions associated to $\sigma$. We can turn on
the auxiliary fields in this multiplet to break the ${\cal N}=2$
supersymmetry to ${\cal N}=1$.\foot{Normally one does not consider
``turning on'' auxiliary fields since their values are fixed
by equations of motion. However, in Wick-rotated signatures (2,2)
or (4,0), there may
be supersymmetric backgrounds which violate equations of motion. For
example, the auxiliary fields $D_{ij}$ in an ${\cal N}=2$
vector multiplet transform
as a triplet under the
R-symmetry group which gets Wick-rotated from $SU(2)$ to
$SL(2)$. For a free ${\cal N}=2$ multiplet, the potential is
$D_{++} D_{--} + (D_{+-})^2$ and one has an ${\cal N}=1$ supersymmetric
background when $D_{++}=-D_{--}=D_{+-}=N$ for any value of $N$.
After Wick-rotation back to Minkowski space, the value of $N$ is
uniquely determined by the reality conditions on $D_{ij}$.
For example, for a free multiplet in Minkowski space, $N=0$ is the unique
supersymmetric background consistent with the reality condition
$D_{++}= (D_{--})^*$. However, in
a non-trivial background such as that of \KlebanovHB\ or \MaldacenaYY,
the reality conditions together with supersymmetry
may imply a non-zero value for $N$.}
For example, we can turn on the
perturbation,
$$ \int d^2z G^- \bar G^- \left[ N (\theta-\bar\theta)^2 \sigma \right]. $$
This corresponds to turning on R-R flux through the cycle
dual to $\omega$, represented by $\epsilon_{\alpha\beta}
\theta^\alpha \bar\theta^\beta ~\sigma$ \VafaWI,
combined with an appropriate amount of 
NS-NS flux, represented by $\epsilon_{\alpha\beta}
(\theta^\alpha\theta^\beta + \bar\theta^\alpha\bar\theta^\beta)
~\sigma$, through the dual cycle. The strength of the NS-NS flux 
(related to the coupling constant $\tau$ of the dual
gauge theory) is dictated
by the condition of extremization of the glueball superpotential \VafaWI,
leading to preservation of ${\cal N}=1$ supersymmetry.
Note that this term reduces the supersymmetry
to ${\cal N}=1$ given by simultaneous shift of $\theta, \bar\theta$.

With these fluxes turned on and the supersymmetry reduced,
the ${\cal N}=2$ vector multiplet is decomposed into
an ${\cal N}=1$ vector multiplet $v_\alpha$ and the
chiral multiplet $t$. These couple to the worldsheet as
\eqn\worldsheetcoupling{
 \int d^2z G^- \bar G^- \left[\left(t + v_\alpha (\theta-\bar\theta)^\alpha
+ N (\theta-\bar\theta)^2 \right) \sigma \right] }
where we included the effect of the fluxes.
In section 2, we saw that, in the $\hat c=3$ model, the
K\"ahler moduli appears as a coefficient of the linear
superpotential \linearpotential .  The coupling \worldsheetcoupling\
in the $\hat c=5$ model can also be written in term of
a superpotential given by
\eqn\cfivesuperpotential{
W = \left( t + v_\alpha (\Theta -\bar\Theta)
 + N(\Theta - \bar\Theta)^2 \right) \Sigma,}
where $\Sigma$ is the superfield in the $\hat c=3$ model
with $\sigma$ as the lowest component, and
$\Theta, \bar\Theta$ are fermionic superfields whose
lowest components are $\theta$ and $\bar\theta$.
The contribution of $W$ to the worldsheet action is
\eqn\Wint{S_{int}= \int d^2 z G^- \bar G^- W. }
Noting that $G^-$ is a linear combination of operators
acting on the 4d part and the Calabi-Yau part, we can express
it as
$$ \eqalign{
\int d^2 z G^- \bar G^- W = &
\int d^2 z \Big[
 \left( t + v_\alpha (\theta -\bar\theta)
 + N(\theta - \bar\theta)^2 \right) G^-_{CY}\bar G^-_{CY}\sigma  \cr
&~~~~+\left(  v_\alpha \p x^{\a \dot -}
 + 2 N (\theta - \bar\theta)_\a \p x^{\a \dot -} \right)
\bar G^-_{CY}\sigma  \cr
 &~~~~\left. +\left(  v_\alpha \bar\p x^{\a \dot -}
 + 2 N (\theta - \bar\theta)_\a \bar\p x^{\a \dot -} \right)
G^-_{CY}\sigma
   - 2 N \e_{\a\b}\p x^{\b\dot -} \bar\p x^{\a \dot -}
\sigma \right] .}$$

Since $W$ is annihilated by
$\oint G^+$ and $\oint\bar G^+$ of \chatgen, $W$ is a chiral superpotential
which implies that
\Wint\ is in the BRST cohomology.
Actually, annihilation by
$G^+$ and $\bar G^+$ of \chatgen\ implies that
$W$ is chiral using the worldsheet equations of motion of the undeformed
theory. In principle, one still needs to check that
$W$ is chiral after including
any possible back-reaction to the worldsheet equations of motion.
Fortunately, there is no back-reaction
to the worldsheet equations of motion for the $d=4$
fields $(x^{\a \dot -},\theta^\a,\bar\theta^\a)$ which appear in $W$.
This is clear since
the equations of motion for these $d=4$ fields come from varying
$(x^{\a\dot +},p_\a,\bar p_\a)$, which are absent from \Wint.

On the other hand, since the vertex operator for the spacetime
curvature and the graviphoton
field strength contain $p_\a$ and $\bar p_\a$,
in the $\hat c=5$ formalism formulated in the last section,
there may be a subtlety
in simultaneously turning on the gravity fields and the
R-R flux. Since it is clear from the target space
point of view that supersymmetry is still preserved with both
of them turned on, there should be a manifestly supersymmetric
description of such a background on the worldsheet.
It would be interesting to understand how to apply the $\hat c=5$
formalism in this case.
On the open string side, turning on the spacetime curvature and
the graviphoton field strength generates the $C$-deformation
of the gluino field \refs{\cdeforma, \cdeformb}. Thus it is
reasonable to expect
a phenomenon dual to it in the closed string side. In the following,
we will consider the large $N$ duality in the absence of the
gravity field strengths.

As in the $\hat c=3$ model for the conifold, the
$\hat c=5$ model has two branches, the $H$ branch with
$\sigma=0$ and the $C$ branch with $\sigma \neq 0$.
We identify each $C$ domain as a hole on the worldsheet.
Whereas the $C$ branch of the $\hat c=3$ model is
described as the Landau-Ginzburg model with the
superpotential \linearpotential\ (and with the
path integral measure $d\sigma/\sigma^2$), the
$C$ branch in the $\hat c=5$ model is the Landau-Ginzburg
model with \cfivesuperpotential . In particular, its
target space is the supermanifold with coordinates
$(\Sigma, \Theta^\alpha, \bar\Theta^\alpha)$.
As in the $\hat c=3$ case, the $C$ branch does
not contribute to a string amplitude unless its
domain has the topology of the disk. This statement
just follows from the functional integral over $\Sigma$
and the operation of $\oint d\sigma \partial/\partial\sigma$
and is independent of whether there are extra degrees
of freedom.

The functional integral over the disk $C$ domain indeed
gives the correct boundary condition for the $N$ D branes
extended in the ${\bf R}^4$ direction with the gluino field
${\cal W}_\alpha$ turned on. To see this, let us
integrate over $\Sigma$ first. As in the case of the
$\hat c=3$ model \OoguriGX, it gives
\eqn\boundarystate{
\oint {d\sigma \over \sigma^2} \exp\left[
-\left( t + v_\alpha (\theta-\bar\theta)^\alpha
 +N(\theta-\bar\theta)^2\right) \sigma\right]
=  t + v_\alpha (\theta-\bar\theta)^\alpha
 +N(\theta-\bar\theta)^2.}
According to the large $N$ duality \VafaWI , $t$ and $v_\alpha$
are related to the open string variable ${\cal W}_\alpha$
as
\eqn\dualitytable{
 \eqalign{ &t={\rm tr} {\cal W}_\alpha {\cal W}^\alpha \cr
&v_\alpha ={\rm tr} {\cal W}_\alpha  \cr
&N={\rm tr} 1 .}}
Using this, the right-hand side of \boundarystate\ can be
written as
\eqn\boundarytwo{
t + v_\alpha (\theta-\bar\theta)^\alpha
 +N(\theta-\bar\theta)^2 = {\rm tr}\left[
\exp\left( {\cal W}^\alpha {\partial \over \partial \theta^\alpha}\right)
(\theta -\bar\theta)^2  \right].}

We can then identify $(\theta-\bar\theta)^2$ as the boundary state
for the D brane extended in the ${\bf R}^4$ direction. As in any state
which is invariant under the topological BRST symmetry, the boundary
state can be decomposed into a chiral primary state and a BRST trivial
part. It was shown in \OoguriCK\ that the chiral primary part is
determined by the (quantum) period of the cycle on which the D brane
is wrapped. For the D brane extended in the ${\bf R}^4$ direction,
the chiral primary part is $(\theta-\bar\theta)^2$; indeed
it imposes the correct boundary condition $\theta^\alpha =
\bar\theta^\alpha$, which is associated with Neumann boundary
conditions for $x^m$. We can then identify the action of
the differential operator
$\exp\left( {\cal W}^\alpha {\partial \over \partial \theta^\alpha}\right)$
as an insertion of $\oint {\cal W}^\alpha (p_\alpha + \bar p_\alpha)$
on the boundary of the disk, giving rise to the correct coupling
of the gluino on the boundary. This shows that the superpotential
for $t$ and the kinetic term for $v_\alpha$ computed in the closed
string theory agree with those for the glueball superfield and
the $U(1)$ part of ${\cal W}_\alpha$ in the open string theory
according to the correspondence \dualitytable . This is what we
wanted to show.

We note that one can start with a different combination of fluxes, for
example,
\eqn\another{ \int d^2z G^- \bar G^-\left[ N
(\theta^1 \pm \bar\theta^1)(\theta^2
\pm \bar\theta^2) \sigma \right], }
and repeat the derivation. (We can also consider more general
quadratic combinations of $\theta$ and $\bar\theta$ that
preserve 4 supercharges. Here we are presenting simple
ones for an illustration.) One will then find the
boundary state whose chiral primary part is represented by
$(\theta^1 \pm \bar\theta^1)(\theta^2
\pm \bar\theta^2)$.
We can interpret it as the boundary
state for a $D_{2n+2}$ brane wrapping on the ${\bf S}^3$ of the
deformed conifold and extending in a $2n$-dimensional plane
in ${\bf R}^4$, where $n$ is the number of minus signs in
\another . This is consistent with what one expects from
T-dual of the open/closed string duality that we discussed in
this paper.

%\MaldacenaRE
\lref\MaldacenaRE{
J.~M.~Maldacena,
``The large $N$ limit of superconformal field theories and supergravity,''
Adv.\ Theor.\ Math.\ Phys.\  {\bf 2}, 231 (1998)
[arXiv:hep-th/9711200].
%%CITATION = HEP-TH 9711200;%%
}

%\PolchinskiMT
\lref\PolchinskiMT{
J.~Polchinski,
``Dirichlet-branes and Ramond-Ramond charges,''
Phys.\ Rev.\ Lett.\  {\bf 75}, 4724 (1995)
[arXiv:hep-th/9510017].
%%CITATION = HEP-TH 9510017;%%
}

The original argument \MaldacenaRE\
for the existence of the large $N$ dualities
of the type discussed in this paper
 starts with the conjectured equivalence of the D
brane description involving open strings and the closed string
description motivated by the computation of the R-R charges
\PolchinskiMT. The result of this paper
provides the worldsheet explanation for the equivalence of
the two descriptions, at the level
of $F$ terms. For the closed
string, the vertex operator $N (\theta-\bar\theta)^2 \sigma$
represents the closed string background with $N$ units
of R-R flux turned on. We have found that turning on this
worldsheet interaction generates
the open string sector whose boundary state for the $4d$ part
of the target space is
represented by $N(\theta-\bar\theta)^2$. This boundary state indeed carries
the correct amount of R-R charge expected from the duality.
We hope that our result in this paper will turn out to be a useful step
toward deriving the full large $N$ duality in the superstring.

\bigskip
\bigskip

\centerline{{\bf Acknowledgments }}

\bigskip

C.V. thanks the hospitality of the theory group at Caltech, where he
was a Gordon Moore Distinguished Scholar. N.B., H.O., and C.V. thank
the Simons Workshop on Mathematics and Physics and the YITP at Stony
Brook for their hospitality during the completion of this work.
H.O. also thanks KITP, Santa Barbara and N.B. thanks Caltech
for their hospitality.

The research of H.O. was supported
in part by DOE grant DE-FG03-92-ER40701.  The research of C.V.
was supported in part by NSF grants PHY-9802709 and DMS-0074329.
The research of N.B. was supported in part by FAPESP grant 99/12763-0,
CNPq grant 300256/94-9 and Pronex grant 66.2002/1998-9.

\listrefs
\end